\documentclass[conference]{IEEEtran}
\IEEEoverridecommandlockouts
\usepackage{cite}
\usepackage{amsmath,amssymb,amsfonts}
\usepackage{algorithmic}
\usepackage{graphicx}
\usepackage{textcomp}
\usepackage{xcolor}
\usepackage{graphicx}
\usepackage{mwe}

\def\BibTeX{{\rm B\kern-.05em{\sc i\kern-.025em b}\kern-.08em
    T\kern-.1667em\lower.7ex\hbox{E}\kern-.125emX}}
\begin{document}

\title{Detection of Distributed Denial of Service Attacks based on Machine Learning Algorithms \\
{\footnotesize \textsuperscript{}}
\thanks{}
}
\author{\IEEEauthorblockN{Md. Abdur Rahman}
\IEEEauthorblockA{\textit{Department of Mathematics} \\
\textit{Jahangirnagar University}\\
Dhaka, Bangladesh \\
marahmanju@juniv.edu}

}
\maketitle
\begin{abstract}
Distributed Denial of Service (DDoS) attacks make the challenges to provide the services
of the data resources to the web clients. In this paper, we concern to study and apply different Machine Learning (ML) techniques to separate the DDoS attack instances from benign
instances. Our experimental results show that forward and backward data bytes of our
dataset are observed more similar for DDoS attacks compared to the data bytes for benign
attempts. This paper uses different machine learning techniques for the detection of the
attacks efficiently in order to make sure the offered services from web servers available. This
results from the proposed approach suggest that 97.1\% of DDoS attacks are successfully
detected by the Support Vector Machine (SVM). These accuracies are better while comparing
to the several existing machine learning approaches. 

\end{abstract}

\section{Introduction} 
Network infrastructures encounters enormous attacks. The Denial-of-Service (DoS) attacks is based on the congestion and one of major threats which break records continuously. These deny the victim to receive services in internet through inundating with malicious traffics.  

These attacks was observed firstly in 1998 [1]. Several well known web sites, such as Amazon, eBay, and Yahoo, etc. were encountered by DDoS attacks in the year of 2000. These web sites were attacked through the internet although these were highly secured web sites to provide services to web clients. These event proves that DDoS attack became a major threat which has to be detected and protected to access data sources and get services by users. The year 2015 and 2016 was the worst year, because DoS attacks was recorded by 500 Gbps and 800 Gbps respectively [2].  

DDoS attacks can be operated on cloud platforms. In this case, attackers use the virtual machines to attack the web site by using VM bots [3]. Attackers rent virtual machines in order to attack because of huge computational ability than using their physical machines.  

Nowadays, there is no debate about the increasing popularity of Internet of Things (IoT), and it is used in the vehicles, wearable devices, and even in home. It need networking to connect public facilities, household appliances, medical equipment, interconnected vehicles, etc.  [4][5][6][7]. One great work used support vector machines (SVM) for the detection of DDoS attacks [8]. Dao et al. (2015) has proposed DDoS attack detection algorithm for SDN devices [9]. Some researcher have proposed about how to boost and improve the IoT security with SDN technology.
One paper proposed a distributed architecture was proposed with security for IoT based SDN domain (Flauzac et al.  [10]). Another paper investigated the potential threats on the Open Flow control channel Li et al.  [11].  
Ahmed and Kim [12] provided guidance for the mitigation of DDoS attacks in IoT. 

Hu et al. (2001) categorized whether the process is normal or intrusive class using  K-nearest Neighbor Classifier [13]. The processes of the same class will make the cluster together. This work had used machine learning exclusively for the detection of attacks. However, this classifier is expensive for computation while the simultaneously increasing the number of processes.   
\begin{figure*}[htbp]
\centerline{\includegraphics[width=1.99\columnwidth]{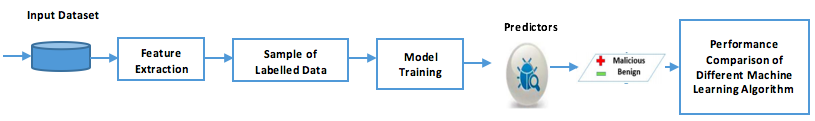}}
\caption{Block diagram of the proposed system.}
\end{figure*}

DOS attack prevents the authentic clients from accessing information from the web server. Two types of DOS attacks are recorded: network level attacks and application level attacks [14]. Network level DoS attacks disable the connectivity of valid users to access network resources, and application level DoS attacks disrupt the services from server resources temporarily or indefinitely. More than 30\% of network attacks are accomplished by DoS attack [15]. These attacks interrupt of accessing to a simple webpage to very large servers. 

\begin{table*}[ht]
\caption{Dataset of Canadian Institute of Cybersecurity} 
\centering 
\begin{tabular}{c c c c c c c c c p{6cm}} 

\hline\hline 

\multicolumn{1}{p{1cm}}{Destination} & \multicolumn{1}{p{1.5cm}}{Flow Duration} & \multicolumn{1}{p{1.5cm}}{Total Fwd Pkts} & \multicolumn{1}{p{1.5cm}}{Total Bwd Pkts} & \multicolumn{1}{p{1.5cm}}{Total Length of Fwd Pkts} & \multicolumn{1}{p{1.5cm}}{Total Length of Bwd Pkts} & \multicolumn{1}{p{1.5cm}}{Initial Window bytes Fwd} & \multicolumn{1}{p{1.5cm}}{Initial Window bytes Bwd} & \multicolumn{1}{p{1cm}}{Label}\\ [0.3ex] 

\hline 
53 & 83718 & 4 & 2 & 184 & 300 & -1 & -1 & BENIGN \\ 
445 & 10706606 & 29 & 24 & 8142 & 4220 & 8192 & 2050 & BENIGN \\
80 & 39723 & 3 & 5 & 26 & 11601 & 8192 & 229 & DDoS \\
443 & 118945 & 19 & 25 & 1169 & 43577 & 29200 & 61 & BENIGN \\
80 & 80803000 & 9 & 6 & 62 & 11607 & 256 & 229 & DDoS \\ [1ex] 

\end{tabular}
\label{table:nonlin} 
\end{table*}

The defense mechanisms for DDoS can be categorized into two main parts: source side defences and destination side defences. It is difficult to recognize the attack from the source side. However, D-WARD [19][20] system was developed to compare incoming traffic with and outcoming traffic on the source side in order to detect DDoS attacks. 

Destination side defense systems can detect and respond the DDoS attacks at the node of victim. Several systems [16] [17] [18] can monitor received packages while detecting the attack. Then it can discard the connection. Meanwhile, network have been vulnerable by attack bundles. As a results, it is much difficult to stop the attack by the attackers.

Several works used four types machine Learning algorithms: unsupervised, supervised,
reinforcement, and semi-supervised learning to train the machine learning models for
evaluating the results [21][22][23]. Moreover, Agrawal et al., 2011 used SVM for the
prediction of total zombies in cyber attacks [24]. One recent work developed an anomalybased application layer of Bio-Inspired for early and fast detection of DDoS attacks from
HTTP flood [29]. Patgiri et Al. have used two machine learning algorithms: Support Vector
Machine and Random Forest, and followed a thorough experiment to detect intrusion. The
performance of these two algorithms is compared to detect intrusion [30]. 

The rest of this paper is structured as follows: In section II, different machine learning techniques are presented for providing basic information. In section III, we present our machine learning framework. In section IV, we discuss simulation results after applying different ML approach, and then compare the results. In the last section, we focus a summary of outcome and future work.

\section{Concepts of Machine Learning} 
We have used different machine learning (ML) approaches such as Logistic Regression (LR), Decision Tree, and SVMs etc. in our system for detection DDoS attack from the benign attempts. These ML approaches are discussed below:
\subsection{Support Vector Machines}
Support vector machines (SVMs) are  the machines that are using to plot training vectors in feature space, and these vectors are labelled by its class. This classification problem is similar to quadratic optimization problem. They uses a technique that is able to avoid “curse of dimensionality”. The predominant feature of SVMs is that it can classify the dataset through determining set (collection) of support vectors made by the set of training inputs, and then a hyperplane is generated in high-dimensional feature
space.  

\subsection{Decision tree (DT)}
Decision tree is used in machine learning for classification. It is efficient way that follows a divide-and-conquer strategy to construct decision tree recursively. The decision tree has the root, internal nodes, branches, and leaves like a tree. Each tree represents a rule which based on the data attributes. Leaves are labeled as the decision for classification. Let the classes are denoted by {$C1$, $C2$,. . ., $C_{n}$}, and each leaf of decision tree is identifying a specific class from class $C_{i}$. 

\subsection{Logistic Regression}
Logistic Regression is one of the most effective classification approaches. It is possible to determine the application layer DDoS attack from the effective features after feature extraction. 
In this paper, we have used logistic regression, however the performance is not suitable for our dataset. 
The logistic regression can be explained as follows: suppose there
are $k$ independent features $x1$, $x2$, .... $x_{k}$, then  the probability of DDoS attack detection is expressed as follows:
\begin{equation}
  P = P(y=1| x_{1}, x_{2}, ... x_{k})
\end{equation}
    
and logistic regression as \begin{equation}
    p = \frac{e^y}{1+e^y}
\end{equation}    
Where, 
\begin{equation}
    y = \gamma_{0} + \gamma_{1}x_{1} + \gamma_{2}x_{2} + .... +\gamma_{k}x_{k},
\end{equation} 
$\gamma_{i}$ is the coefficient, and x1,$x1$, $x2$, .... $x_{k}$ are the features.

\section{Proposed Framework}
We observe datasets in which objects in a specific group are related to each other, and different from objects of other groups. As our datasets contains two types of data: DDoS attack and benign. As our goal is to observe anomalies, we observe if there are any data which is different from normal data. The advantage of using machining learning techniques is that they can separate DDoS attack objects from datasets from the benign objects. 

\begin{figure*}[htbp]
\centerline{\includegraphics[width=1.99\columnwidth]{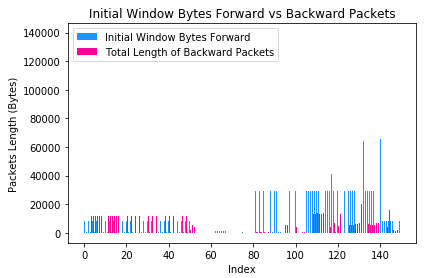}}
\caption{Forward and Backward packets indicates the Data for DDoS attacks and Benign Attempts.}
\end{figure*}

This framework contains of several components: feature selection, data pre-processing, data analysis for different machine learning algorithms, training the dataset to algorithms, testing the dataset and then comparing the results with different algorithms. The block diagram of the proposed system is shown in Fig.
1. The machine learning DDoS attack detection system consists of the following phases:

Preprocessing: it is required to make collected data to an understandable format so that it must be complete, consistent and free from lacking in specific behavior.  

Training: several machine learning algorithms is trained using the machine readable data. The predominant thing is that our machines experiences through these data which have DDoS attacks and normal data. The features of these training data fall into two classes: benign and attack.

Testing: after machine learns from the training dataset, then it can make predictions from new dataset based on its learning. This is for measuring the performance on testing data.   

We have developed a classifier that can classify malicious packet from the benign packet. This model works as detectors which firstly detect it and then stop or minimize the strength of an attack. Indeed, this detector receives the request from the web clients, then it can identify the malicious packet if this request is falling into the DDoS class. This is detected as this request does not behave normally. 
In this paper, we focus on different Machine Learning Algorithms such as Support Vector Machine (SVM), decision tree, and logistic regression. This framework offers  robust techniques to the DDoS attack detection.  

Also, our classifier identifies irregularities in the network. The predominant thing is that this classifier has to permit authentic packets for passing through the network so that these must be reached to the destination without any interruption or delays. For providing the services to the legitimate clients, this detectors must check each request precisely.

After loading datasets using pandas, we have chosen two attributes, initial window forward bytes and backward bytes, to observe the main trends of attacks and benign attempts. The following figure presents a visualization of the relation between these attributes (Fig. 2) in which the bar diagram for the index 1 to 50 have simillar patterns, and the rest of bars have much fluctuations. After analysing data, we observed the DDoS attacks data represented in index until 50.   

We propose a DDoS attack detection system in order to detect attacks, and then mitigate the impact of the attacks. We analyze the detection accuracy for DDoS attack using different machine learning techniques.

\section{Simulation Results} 

We work on the client of 1.6 GHz Intel Core i5 with macOS High Sierra for implementing the different machine learning algorithms to detect DDoS attacks using Python  (Scikit-learn library). 

\begin{figure*}[htbp]
\centerline{\includegraphics[width=1.99\columnwidth]{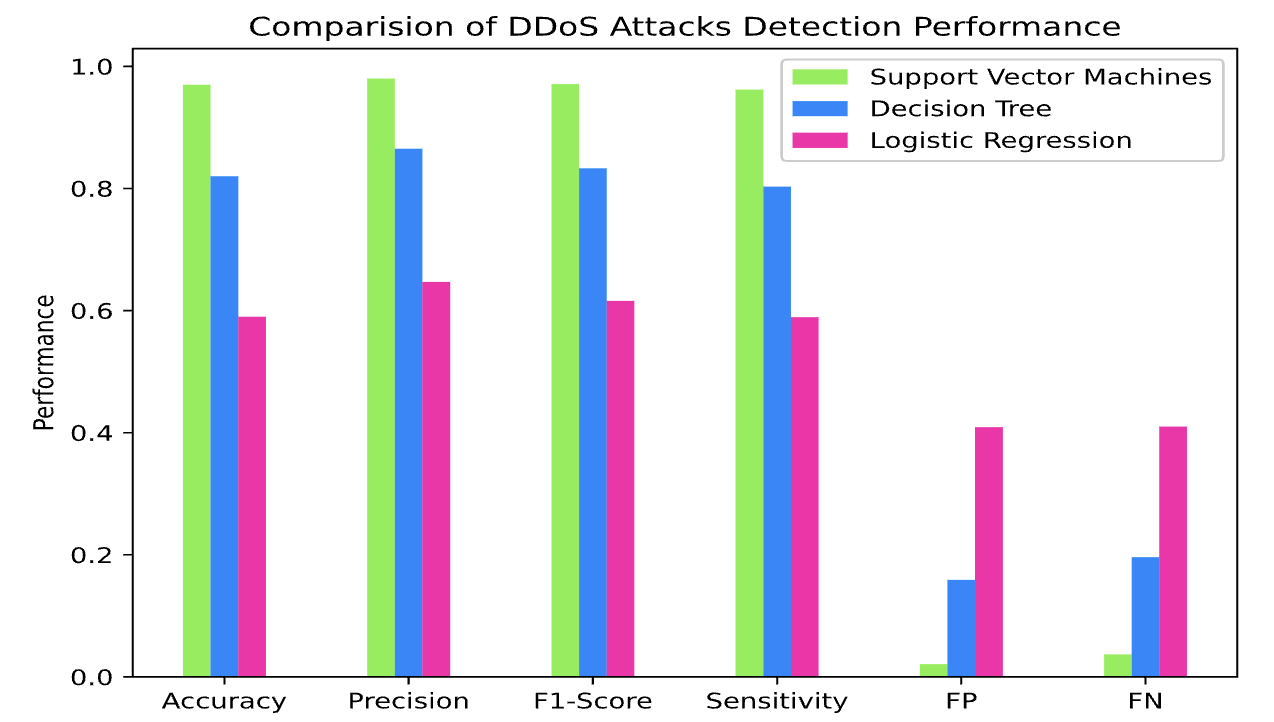}}
\caption{A Bar Diagram to Show Performance of Three Machine Learning Algorithms.}
\end{figure*}

\begin{table}[ht]
    \centering
\caption{PERFORMANCE OF MACHINE LEARNING ALGORITHMS} 
\centering 
\begin{tabular}{c c c c c c c} 
\hline
\textbf{Method} & Accu. & Precision & Recall & F1-Score & FP & FN \\ [0.2ex] 
\hline 
\\
Logistic\\ Regression & 0.593 & 0.647 & 0.616 & 0.589 & 0.409 & 0.410 \\ \hline
Decision \\ Tree       & 0.827 & 0.865 & 0.833 & 0.803 & 0.159 & 0.196 \\ \hline
SVM                 & 0.971 & 0.980 & 0.971 & 0.962 & 0.021 & 0.037 \\ \hline
    \end{tabular}
\end{table}

\subsection{Data Collection}
In the simulation, we uses the dataset of Canadian Institute of Cybersecurity which includes the two types of objects: BENIGN and DDoS attack. This attack is accomplished over various network and sessions. These sessions became to attack and non-attack phases. Table I represents the sample of dataset for this proposed work.  

We examine the features of the dataset and observe nine parameters. However, four attributes of them have a good indication for the detection of DDoS attack. Some records of datasets have some similarities. That means that intruder sends the data which have little variation while requests from different web clients have much variations as legitimate web clients are different users, and they request servers for different requirements. We observe that their data has huge variations than attackers shown in Fig. 2.  

The plot shows the little variations of data in the first part which indicates very suspicious known as DDoS attack in which the attacker sends the almost same length of forwarding data for attacking the servers shown in Fig. 2. On the other hand, the data values of features are very distinguishable in the second part of the data. It has great variations of data in the second part which is the indication of normal data. This is because different web clients request to the servers for varieties of resources. That is the main reason to have huge variations of data.

\begin{figure*}[htbp]
\centerline{\includegraphics[width=1.2\columnwidth]{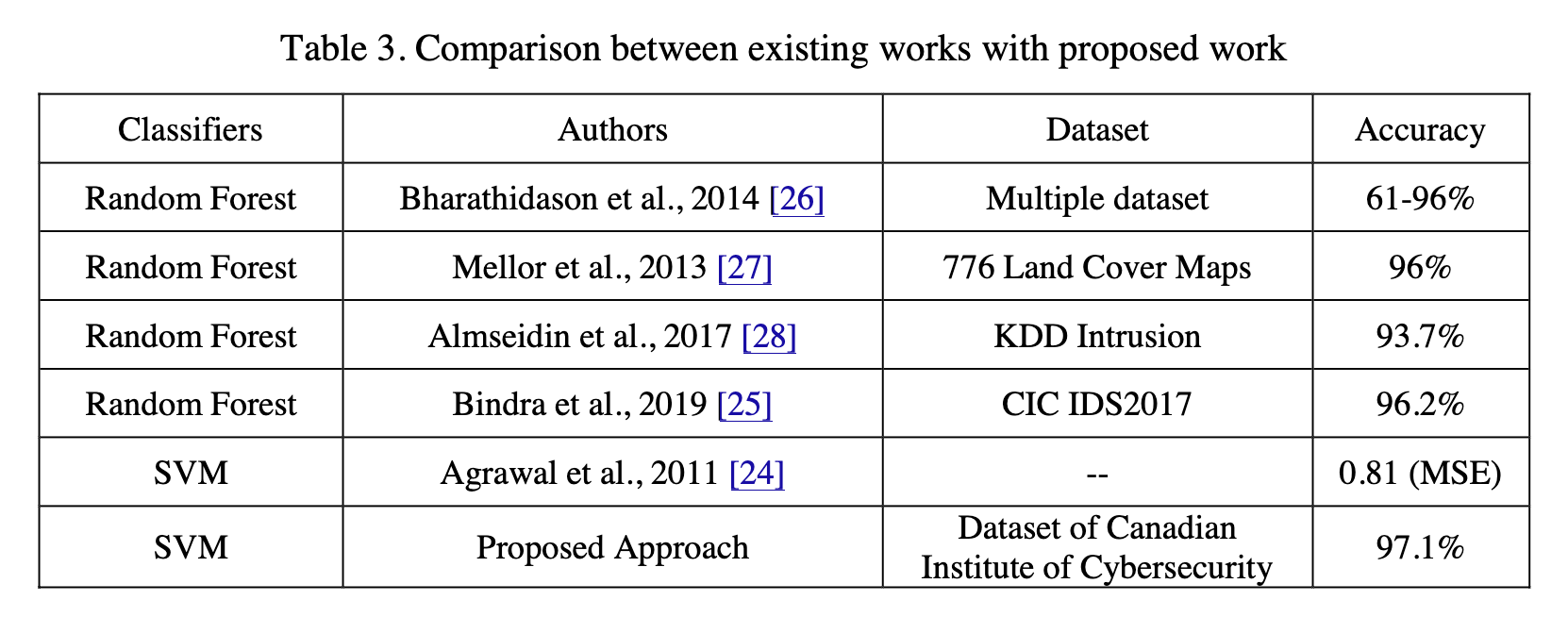}}

\end{figure*}

\subsection{Prediction Accuracy}
This model can make the predictions based on the data for which the correct labels are
assigned. We observed categorical data in the label so that each object is identified whether
this comes from the web clients request or attacker.
It is important to know what different types of data are sent by the web clients and
attackers. At the same time, there should be data which is responded by the web servers. We
have examined closely about the differences among these two categories objects. From the
dataset, the selected data of objects can be set into a NumPy array in python. The dataset is
divided into training set and the test set. This training data have been feed to the proposed
model which is SVM(s) classifiers. Using this data, this model learns about the situations.
Then, it can predict for each test data for comparing it against its label. The accuracy for
prediction correctly is measured which expresses how fine this model works. For this
classifier, the accuracy of test set is almost 0.971, which indicates that the percentage of
prediction is 97.1\%. According to the mathematical expectations, our model meets the 97.1\%
correct through its forecasts.

\subsection{Evaluation through confusion matrix}
It is imperative to keep in remember that misclassifying a DDoS attack makes the negative
impression of the model, although it can classify the benign attempt correctly. We focus to
calculate the true positives and true negatives for the confusion matrix to know the success or
failure of this work.
In terms of detection accuracy, training time, running time, scalability, Support Vector
Machines (SVMs) achieve expected outcome and beat other techniques like decision tree,
logistic regression, etc. while testing three different Machine Learning classifiers on the
dataset. More importantly, we observed that SVMs have high detection accuracy of the DDoS
attack among them, which is 97.1\%. But, minimum accuracy is recorded for logistic
regression algorithm (59.3\%). [Table 2] represents the performance score board of SVM,
decision tree, and logistic regression. [Figure 3]. represents a bar diagram to show
performance of three machine learning algorithms.

We have compared the proposed approach with some exiting works in which our approach
has high detection accuracy of the cyber attack while comparing random forest classifiers. As
our accuracy is 97.1\% which is higher than other classifiers expressed in Table 3. The closest
accuracy is observed in the work of Mellor el al. [27] which is 96\%. Moreover, Agarwal et al.
[24] used SVM and it has MSE score 0.81.
The true positive rate is the attempts of DDoS attacks correctly identified by the algorithm.
classified correctly, and true negative rate is not the attempts of DDoS attacks correctly
identified by the algorithm. Also, the false positive rate is the proportion of benign attempts
classified as DDoS attacks, and false negative rate is similar to the proportion of DDoS
attacks classified as benign attempts.

Let us denote the number of benign attempts classified as benign as $benign_{benign}$, the number of benign attempts classified as $DDoS$ attacks as $benign_{DDoS}$, the number of $DDoS$ attacks classified as benign as $DDoS_{benign}$, and the number of $DDoS$ attacks classified as $DDoS$ attacks as $DDoS_{DDoS}$. We then define $fp$, the false positive rate, as 

\begin{equation}
  fp =
    \frac{benign_{DDoS}}{{benign_{DDoS}}+{benign_{benign}}}
 \end{equation}
\hspace{.3cm}  and $fn$, the false negatives rate, as 
 \begin{equation}
  fn =
    \frac{DDoS_{benign}}{{DDoS_{benign}}+{DDoS_{DDoS}}}
 \end{equation}

Following this definition, $fp$ = 0.009 will correspond to one of every 100 benign attempts being classified as DDoS, and $fn$ = 0.045 would correspond to five of every 100 DDoS attacks being classified as benign attemps. These terms $fp$ and $fn$ which are used in this work for showing the evaluations. 

\section{CONCLUSION}
We have applied different machine learning algorithms for detecting the patterns of DDoS
attacks. We also validate their performance for ranking the best ML algorithms for serving
these purposes. In terms of detection accuracy, training time, running time, scalability,
support vector machines (SVMs) achieve expected outcome and beat other techniques like
decision tree, logistic regression, etc. while testing three different Machine Learning
classifiers on the dataset. More importantly, we observed that SVMs have high detection
accuracy of the DDoS attack among them, which is 97.1\%. But, minimum accuracy is
recorded for the logistic regression algorithm (59.3\%). The range of the accuracy of our three
classifiers is approximately 59.3\% to 97.1\%. These results encourage to do additional
research for the detection of DDoS attack to protect servers to serve their assigned services.
This work has some imitations. We will address to solve this issue in our next work. 
\\
This work has some limitations. We will address to solve this issue in our next work.

\section*{Acknowledgment}

\vspace{12pt}
\color{red}

\end{document}